\documentclass{article}
\def\abstract#1{\vskip 7mm
        \begin{center}{\large Abstract}\par \smallskip
                \begin{minipage}[c]{12cm}
                        \small #1
                \end{minipage}
        \end{center}
}
\def\title#1{\begin{center}{\Large\bf #1}\end{center}}
\def\author#1{\vskip 5mm \begin{center}{#1}\end{center}}
\def\address#1{\begin{center}{\it #1}\end{center}}
\makeatletter
\@ifundefined{lesssim}{\def\lesssim{\mathrel{\mathpalette\vereq<}}}{}
\@ifundefined{gtrsim}{\def\gtrsim{\mathrel{\mathpalette\vereq>}}}{}
\def\vereq#1#2{\lower3pt\vbox{\baselineskip1.5pt \lineskip1.5pt
\ialign{$\m@th#1\hfill##\hfil$\crcr#2\crcr\sim\crcr}}}
\makeatother

\begin{document}

\title{%
Black Hole Production at a TeV\footnote{Invited talk at the 12th
Workshop on General Relativity and Gravitation, Komaba Campus, University
of Tokyo, Tokyo, Japan November 25-28, 2002.}
}
\author{%
  Roberto
Emparan\footnote{E-mail: emparan@ffn.ub.es}
 }
\address{%
  Theory Division, CERN,
  CH-1211 Geneva 23, Switzerland\\and\\
ICREA and Departament de F{\'\i}sica Fonamental, \\Universitat de Barcelona,
Diagonal 647, E-08028 Barcelona, Spain\footnote{Permanent address from
January 2003.}
}

\abstract{
In this talk I briefly discuss some features of a striking signature of
TeV quantum gravity/strings and large extra dimensions: the production
of exotic objects such as black holes, and their stringy precursors, the
string resonances at large masses which can be dubbed `string balls'.
The focus is on aspects less frequently discussed in other reviews of
this subject.
}

\section{Introduction}

Let us start by asking one of the most basic questions in theoretical
physics: What is the fundamental scale for Quantum Gravity? Answer:{\it
We don't know.} It is only quite recently that we have become fully
aware of this. The reason for our ignorance is quite simple: we do not
know how to formulate the correct theory of quantum gravity. What we do
know is the following: the fundamental energy scale $M_*$ has to be
larger than a TeV, for otherwise we should have seen effects of quantum
gravity at LEP and the Tevatron\footnote{There are ways of sidestepping
even this lower bound \cite{dgetc}, but I will not discuss these
scenarios here.}. We also know that $M_*$ can be as large as, but not
larger than, the familiar Planck scale $M_{(4)}\sim 10^{19}$ GeV, since
at this energy four-dimensional gravity becomes strongly coupled and
quantum effects surely must become important. The phrase {\it
four-dimensional} in the last statement betrays already the mechanism
that allows $M_*$ to be smaller than $M_{(4)}$ \cite{add}. The introduction of large
extra dimensions dilutes gravity in an internal
$n$-dimensional volume $V_n=L^n$, and $M_{(4)}$ is revealed as an effective
coupling 
\begin{equation} 
M_{(4)}^2=M_*^{n+2}L^n\,,
\end{equation}
describing the strength of gravity only at distances larger than the
length scale of the additional dimensions.

The scenario is completed by a mechanism that prevents all particles and
interactions, other than gravity, to spread out into these large extra
dimensions: they are confined inside a brane, of thickness $\lesssim$
TeV$^{-1}$, localized in the internal space. By making $L^n$
sufficiently large to allow $M_*$ to be near the TeV scale, one achieves
a reformulation of the hierarchy problem.

The great interest in these models has been prompted by the fact that,
with $M_*$ as low as a TeV, one obtains spectacular new physics in future
colliders such as the LHC. The signatures due to graviton production and
exchange have been thoroughly studied (see e.g. \cite{rpp}). In this
talk I will briefly review some aspects of a more striking signature of
TeV quantum gravity/strings and large extra dimensions: the production
of exotic objects such as black holes
\cite{argy,bafi,ehm3,gith,dimland}, and their stringy precursors, the
string resonances at large masses which can be dubbed `string balls'
\cite{strball}. Both these objects have in common their large mass
(larger than fundamental mass scale), and the fact that they have a
large degeneracy of internal states. The latter is crucial to their
large production cross section.

There are already a number of reviews on this subject \cite{reviews} to
which the reader is referred to for other details, for discussion of
topics that I mentioned in my talk but are not included here such as
black hole production in cosmic rays \cite{cosmic}, and for extensive
lists of references. Instead, I will mostly focus on aspects that have
been less prominent in former discussions in the literature.

\section{Black holes on branes}

Finding the exact description of a black hole localized on a
three-brane, whether the bulk spacetime is compact (as in ADD scenarios)
or non-compact (as in the RS2 model), is an outstanding problem in
General Relativity. However, the problem can be sidestepped for a black
hole formed at a typical collider energy in the range of $O(10)$ TeV.
Such a black hole has a size $\sim 10^{-15}$ mm, which is several orders
of magnitude smaller than the size that the extra dimensions can
presumably have (e.g., for $n=7$ and $M_*\sim O($TeV$)$ we would have
$L\sim$ fermi). Therefore, at distances smaller than $L$ we can neglect
all finite-size effects from the compact dimensions, and approximate the
metric for a neutral, non-rotating black hole by the
Tangherlini-Schwarzschild solution in $4+n$ dimensions. We want this
black hole to be stuck to the brane, in such a manner that, at distances
$r\ll L$ the metric induced on the brane is
\begin{equation}
ds^2_{brane}\simeq -\left(1-\frac{\mu}{r^{1+n}}\right)
dt^2+\frac{dr^2}{\left(1-\frac{\mu}{r^{1+n}}\right)}+r^2
d\Omega^2_{(2)}\,.
\end{equation}
All the interesting physics (the formation and evaporation of the black
hole) happens in a region within a few horizon radii, so this
approximate geometry will be enough for our purposes. Note that
this is not a solution of the Einstein equations in four dimensions: the
additional modes of gravity in the bulk (Kaluza-Klein modes) are not
zero in the region near the black hole horizon. It can also be argued that
the mass of the black hole as measured from the $4+n$ dimensional point
of view is the same as the one that a four-dimensional observer,
probing distances $r\gg L$, would measure \cite{ehm3}.

\section{Scattering at $E> M_*$ and Black Hole Formation}

Given our ignorance about the correct theory of quantum gravity, one
might think that we cannot say anything at all about scattering at
energies beyond the fundamental Planck scale $M_*$. This is not quite
correct.\footnote{Ideas similar or related to those in this section have
been discussed previously in, e.g.,
\cite{thooft,acv,bafi,eshock,gith,grw}.} While there is certainly a gap
in our ability to perform calculations at energies near the fundamental
scale of quantum gravity, $E\sim M_*$, a number of generic arguments
based on a crucial property of General Relativity, namely the appearance
of horizons, allow us to argue for a description of some features of
scattering in the transplanckian energy regime. There are certainly
effects that, at present, we do not even have a proper theory to
calculate, but in many instances an asymptotic observer will be causally
cut off from them. The crucial point is cosmic censorship, which can be
taken to say, quite liberally, that the extreme high energy regime will
be shrouded by a horizon. Two particles scattering at transplanckian
center-of-mass energy, $\sqrt{s}$, and small impact parameter $b$, will
produce a large concentration of energy in a small volume, and therefore
will induce a large spacetime curvature, larger than the fundamental
scale. To a low-energy observer, this will be equivalent to a curvature
singularity. Then, if cosmic censorship holds, a horizon will appear,
hiding these higher-energy effects from the observers outside. 

This generic argument assumes that in the collision of two
transplanckian particles their energy actually gets concentrated in a
small volume. One may worry that the emission of a large fraction of
energy into initial state radiation might prevent this. However, this
does not seem to be the case. The analyses of \cite{shocks} show that
trapped surfaces do form in the collision of classical transplanckian
particles. Quite generically, these trapped surfaces appear before the
colliding particles are at a distance where quantum gravity corrections
(e.g. higher-dimension operators in the low energy effective action)
must be taken into account. Once the trapped surfaces arise, a
singularity will inevitably develop, and a horizon is expected to censor
it out. The area of the apparent horizon (the outer envelope of the
trapped surfaces), which cannot be larger than the area of the event
horizon, is found to be smaller than, but still close to the area of a black
hole of mass equal to $\sqrt{s}$. Hence follows a (dimension-dependent)
upper bound on the fraction of gravitational radiation that is
classically produced.\footnote{Still, the question remains of whether
the radiation of quantum origin is also small. I thank Riccardo Rattazzi
for discussions on this issue.} 

Low energy effective field theory predicts the failure of General
Relativity at energies near or above the Planck scale by the effect of
higher-dimension operators in the low energy effective action becoming
as important as the Einstein-Hilbert term. What the argument above says
is that the effects of these operators will be largely confined inside a
black hole horizon. An alternative view, using a spacetime description,
would be the following: a region of large curvature can be enclosed by a
surface where all curvatures are moderate; say, below a given cutoff
curvature radius. This region can be cutoff and then replaced by
boundary conditions at its surface. The main assumption, analogous to
the locality of counterterms in the low energy effective action, is that
the specification of these boundary conditions is sufficient to encode
all the effects that the physics at higher curvatures would have on the
lower curvature regions. The situation changes if a horizon appears,
since it not only excises the large curvature region, but the horizon
itself also provides a set of boundary conditions (i.e., thermal) that
is largely independent of the details of the large curvatures inside the
horizon\footnote{The extent to which the properties of the horizon are
so universal and independent of the physics at short distances is
closely related to the black hole no-hair conjecture, and to the problem
of information loss.}.

These considerations reveal an essential property of quantum gravity:
its softness at high energies. The higher the energy of the collision,
the larger the horizon that forms, and therefore the less important that
the quantum gravity corrections will be for an observer outside the
horizon. So gravitational scattering at very high energies becomes
softer and softer. Ultimately, the outcome of the scattering will be the
Hawking radiation emitted in the decay of the black hole, whose
wavelength is comparable to the black hole size, and therefore larger
for higher energy collisions. This softness is also a property that can
be observed in the eikonalized gravitational scattering amplitudes, in
which one can have a large momentum transfer, but built out of a very
large number of small momentum transfers, so at the end one does not
probe the very shortest distances \cite{thooft,acv}. It is also strongly
reminiscent of the softness of strings in the regime of deep inelastic
scattering. The reasons for it are at first sight quite different. One
can see the parallels both in the Regge regime and in the deep inelastic
scattering regime, and in fact the black hole/string correspondence
principle suggest such a connection, which to our knowledge has not been
discussed in detail in the literature. The correspondence principle
also allows us to use string theory to perform a simple calculation
that leads to the correct estimate for the black hole production cross
section. This will be discussed in Sec.\ 5 below.

At transplanckian energies
one must go beyond perturbation theory, but it is the {\it classical}
gravitational dynamics that dominates the evolution of the
system\footnote{Semiclassical, if one also includes the decay via
Hawking radiation.}. This is apparent both in the case where a black
hole forms, and also in the eikonal analysis of the scattering. It can
be interpreted as saying that one must consider diagrams with an
arbitrary number of vertex insertions, but where graviton loops are
absent; matter lines in the diagrams are taken to be on-shell.

A full calculation of the production cross section, using classical
General Relativity, is a very complicated task. A simple estimate can be
obtained by taking into account the above-mentioned bounds on initial
state radiation, and other factors, such as the fact that the capture
cross section of a black hole is actually larger than its area, which
may raise the cross section. For practical purposes, it seems reasonable
to estimate the black hole production cross section as the geometric
cross section
\begin{equation}\label{sigmabh}
\sigma\simeq\pi R_H^2 \sim
\frac{1}{M_*}\left(\frac{\sqrt{s}}{M_*}\right)^{\frac{2}{n+1}}\,.
\end{equation}
Since this cross section does not contain any small numbers
that might suppress it, at energies above $M_*\sim$ TeV it leads to
enormous cross sections, on the order of TeV$^{-2}\sim O(100)$ pb. More
careful estimates result in production rates at the LHC as large as a
black hole per second \cite{dimland,gith}.

\section{Black hole decay}

Once the black hole forms, it will decay very rapidly due to Hawking
evaporation. But, where is this radiation emitted? This is a crucial
point, since if the radiation were emitted into the KK modes of the bulk
graviton it would escape our detectors and we would not be able to
distinguish black hole formation from other missing energy events. In
fact, a simple argument appears to lead precisely to this conclusion:
Since the extra dimensions are assumed to be much larger than the black
hole, from the point of view of a four-dimensional observer there is an
enormous number of light KK modes, of mass smaller than the Hawking
temperature. Emission into these modes would then overwhelm the
radiation of the Standard Model particles that are emitted along the
brane. Viewed another way, the phase space available for emission into
the bulk is much larger than for emission along the brane, making
virtually zero the fraction of energy that could possibly be detected.

Even if it was clear from early on that TeV-gravity scenarios greatly
enhance black hole production \cite{argy,bafi}, these arguments
prevented further discussion of their possible detection at colliders.
However, in \cite{ehm3} a number of counterarguments were given to
conclude that the outcome of the evaporation is emitted mostly along the
brane and therefore can actually be detected. This opened the door to
the studies of the actual phenomenology of the process
\cite{gith,dimland}. 

Instead of repeating the arguments of
\cite{ehm3}, here I will present an ellaboration of an equivalent but
slightly different argument, originally due to L.\ Susskind.

The (erroneous) argument based on the fact that the number of KK modes
is much larger than the number of brane modes assumes that Hawking
radiation is emitted with roughly the same probability into all these KK
modes and into the brane modes. Let us revisit this line of reasoning,
but instead of labelling the KK modes by the momentum in the extra
dimensions, in a plane-wave basis decomposition, we will decompose the
modes in a basis of spherical harmonics. Then we classify the modes
according to their partial-wave number $\ell$. Viewed this way, the fact
that there are many more modes coming from the $4+n$-dimensional bulk
than from the four-dimensional brane is reflected in the fact that, for
large $\ell$, the degeneracy of a mode of given $\ell$ grows in the
former case as $\propto\ell^{n+1}$, much faster than the degeneracy on
the four-dimensional brane, which is well-known to be $2\ell+1$, i.e.,
$\propto \ell$. Imagine now a hot object of characteristic size $R$,
which is radiating thermally at a typical wavelength $\lambda\ll R$.
There will be radiation into modes of partial-wave number up to the
large value $\ell_{max}\sim R/\lambda$. Since there are many more of
these modes in higher dimensions, much more radiation will go into the
bulk than into the brane. 

However, a black hole is not this sort of object. The typical wavelength
of Hawking radiation is actually on the order of, or a little larger
than, the black hole size, $\lambda_H \gtrsim R_{bh}$. So the black hole
emits predominantly into s-wave modes with $\ell=0$. The black hole does
not make use of the large number of highly degenerate $\ell$-angular
modes that the higher-dimensional bulk provides. Since s-waves are
non-degenerate, their number of modes is the same whether the field
propagates in four or $4+n$ dimensions, so each brane mode is emitted at
roughly the same rate as a bulk mode. In other words, the black hole
does not see the angular directions, only the radial direction, which is
common to modes propagating in any number of dimensions. In typical
brane-world scenarios there are $\sim 60$ Standard Model modes
propagating along the brane, versus a single graviton propagating in the
bulk. Then the black hole will radiate mainly on the brane.

The evaporation time for one of these black holes will be extremely
short. The black hole will typically be only a factor $\sim O(10)$
larger than the TeV scale, so its lifetime will be $\tau \sim O(10)$
TeV$^{-1}\sim 10^{-27}$~s, much shorter-lived than a typical hadronic
resonance. The collider signatures for this Hawking radiation were
analyzed in \cite{gith,dimland}. One should look for events with high
multiplicity, due to the large number of quanta emitted during the
evaporation. Events could be tagged by the presence of promt leptons and
photons, of energy $\sim O(100)$ GeV, and of similarly very energetic
jets.

\section{String balls}

The semiclassical description of a black hole must break down when its
mass is close to the fundamental mass scale of gravity $M_*$. But in
string theory, if the string coupling $g_s$ is weak, this description
breaks down even at higher masses. When the black hole radius $R_H$ gets
smaller than the string length $\ell_s$ (and the black hole temperature
gets higher than the Hagedorn temperature), the string corrections to
the low energy effective action cannot be argued, as we did before, to
be hidden inside the horizon, so one must resort to a stringy
description. The critical mass where this happens, i.e., the minimum
mass for which a semiclassical description of the black hole is
reliable, is
\begin{equation}\label{corrmass} 
M_{min}\sim M_s/g_s^2\,,
\end{equation}
and, for small $g_s$, this can be quite larger than $M_*\sim g_s^{-2/n+2}
M_s$. In particular, this means that if $g_s\lesssim 0.3$, the LHC may
not be able to probe the regime of black holes.

In this case, what one would be producing, instead of a black hole, is a
highly excited configuration, a long, jagged, very massive string
resonance: a {\it string ball}\footnote{The phrase ``ball of string''
had been used before in a slightly different context in \cite{bogi}.}.
Over the years, a concept of correspondence between black holes and
strings has been developed \cite{corr}, according to which a black hole
that evaporates down to the mass (\ref{corrmass}) makes a transition
into a string ball. Subsequently, the string ball evaporates by emitting
massless string modes at the Hagedorn temperature.

The production cross section for a string ball can be computed using
string perturbation theory, by factorizing the four-point amplitude for
forward, elastic string scattering in the resonant $s$-channel --
one of the earliest calculations in string theory. This yields
\begin{equation}
\sigma \sim g_s^2 s/M_s^4\,.
\end{equation}
When viewed in the $t$ channel, this four-point amplitude is dominated
by graviton exchange. The latter indeed determines both the factor
$g_s^2$, from tree-level exchange of a closed-string state, and the
energy dependence $\propto s$, from the graviton pole. Therefore the
result is universal for all string theories, since they all contain a
graviton. So, besides factors that depend on the polarization of the
ingoing particles, the result is independent of the initial states, as
long as they are much lighter than $\sqrt{s}$.

This perturbative calculation will receive corrections at energies where
the unitarity bounds are saturated, $g_s^2 s/M_s^2 \sim 1$. Above these
energies, it seems reasonable to assume that the cross section saturates
and remains constant at $\sigma \sim l_s^2 \sim M_s^{-2}$. This cross
section then extends up to the energies (\ref{corrmass}) where the black
holes form, and interpolates between all these regimes in a
parametrically smooth manner. This simple argument provides
evidence in favor of the geometric cross section (\ref{sigmabh}).

It is clear that in scenarios with string scale near the TeV, string
balls will be produced at rates as large as the black hole production
rates discussed above. In fact, it may happen that only the
string ball regime is accesible to the LHC. The decay modes and
properties of string balls are also quite similar to those of black
holes. A main difference is that the temperature of the radiation
emitted (the average energy of the decay products) is fixed at the
Hagedorn temperature, and is essentially independent of the mass of the
string ball, whereas for black holes the temperature is smaller for
larger black holes. This may help distinguish between both kinds of
objects, and may also allow for a determination of parameters such as
the string mass $M_s$.

%\section{Cosmic rays}

\bigskip

{\bf Acknowledgements} 

I would like to thank the organizers of the XIIth
Japanese General Relativity and Gravitation Meeting, and in particular
Tetsuya Shiromizu, for the invitation to such a delightful conference.

\end{document}